\documentclass[conference, draftclsnofoot, onecolumn]{IEEEtran}
\usepackage{cite}
\usepackage[pdftex]{graphicx}
\usepackage{amsmath}
\usepackage{subcaption}
\usepackage{amssymb}
\usepackage[ruled, linesnumbered, commentsnumbered, longend]{algorithm2e}
\usepackage{floatrow}
\usepackage{xcolor}
\usepackage{optidef}

\newcommand{\ie}{i.e.~}

\begin{document}
\bstctlcite{IEEEexample:BSTcontrol}
\title{Mean-Field Game and Reinforcement Learning MEC Resource Provisioning for SFC }

\author{
\IEEEauthorblockN{Amine Abouaomar\IEEEauthorrefmark{1}\IEEEauthorrefmark{3}, Soumaya Cherkaoui\IEEEauthorrefmark{1}, Zoubeir Mlika \IEEEauthorrefmark{1}, and Abdellatif Kobbane\IEEEauthorrefmark{3}}
\IEEEauthorblockA{\IEEEauthorrefmark{1}INTERLAB, Engineering Faculty, Université de Sherbrooke, Sherbrooke (QC), Canada.}
\IEEEauthorblockA{\IEEEauthorrefmark{3}ENSIAS, Mohammed V University, Rabat, Morocco.}
}

\maketitle

\begin{abstract}
In this paper, we address the resource provisioning problem for service function chaining (SFC) in terms of the placement and chaining of virtual network functions (VNFs) within a multi-access edge computing (MEC) infrastructure to reduce service delay. We consider the VNFs as the main entities of the system and propose a mean-field game (MFG) framework to model their behavior for their placement and chaining. Then, to achieve the optimal resource provisioning policy without considering the system control parameters, we reduce the proposed MFG to a Markov decision process (MDP). In this way, we leverage reinforcement learning with an actor-critic approach for MEC nodes to learn complex placement and chaining policies. Simulation results show that our proposed approach outperforms benchmark state-of-the-art approaches.
\end{abstract}


\IEEEpeerreviewmaketitle

\section{Introduction}
Future wireless networks, including the fifth-generation (5G), are being deployed worldwide giving birth to a whole new generation of services with stringent quality-of-service (QoS) and quality-of-experience (QoE) \cite{herrera2016resource}. To reach their performance targets, future generations of wireless networks rely on multiple innovative technologies such as the multi-access edge computing (MEC)  \cite{filali2020multi,9318243}, the network function virtualization (NFV) \cite{laaziz2019fastscale}, and the software-defined networks (SDN)\cite{alam2020survey, 8004158} . A service can be seen as a set of network functions working sequentially to deliver the requested service. These services rely on dedicated hardware that often fails to keep up with the evolution of customer's requirements \cite{yang2020recent, 7585028}. To cope with such a challenge, network functions are now virtualized within virtualization infrastructures such as clouds or MEC nodes, instead of being physically placed and configured. However, this approach requires efficient resource allocation schemes for the infrastructure.

To offer diverse services efficiently, service providers tend to combine sequences of network functions or even linking existing services with each other. This new form of services is referred to as service function chaining (SFC). SFCs will for sure reduce cost and the capital investment in dedicated hardware and will help reduce the latency in cases their deployment is done at the edge level in MEC nodes. Such requirements open the door for new challenges related essentially to the SFC resource provisioning at the edge of the network \cite{kaur2020comprehensive}. The SFC resource provisioning takes many shapes, which can be seen as a problem of placing the VNFs into adequate places to reduce the cost in terms of investment, latency, and resource consumption. Another point of view suggests that the SFC resource provisioning is a problem of chaining the VNFs. And other works consider it as the problem of scheduling the execution sequence of the VNFs. These points of view can be grouped into two perspectives, the VNFs placement, and chaining, and the SFCs scheduling \cite{herrera2016resource, abouaomar2021service}. In this paper, we are more interested in tackling the problem of VNFs placement and chaining within a MEC beyond 5G networks. The VNFs placement and chaining consist in finding an adequate scheme to place the VNFs within a MEC node regarding the resource state of the MEC nodes themselves, and the capacities of the links between the different MEC nodes. The VNFs placement and chaining problem in literature is studied using many paradigms, such as machine learning \cite{8647858}, integer programming \cite{9013429}, and meta-heuristics approaches \cite{9062531}. However, these works, when solving the SFC resource provisioning problem through VNFs placement and chaining, did not consider some important aspects such as the dynamic of the request arrivals, the request heterogeneity, and the type of resources under study.

In this paper, we propose a theoretical game study for the SFC resource provisioning within a MEC context in 5G and beyond networks. As opposed to previous works, we assume that the SFCs' demands, in terms of resources, are heterogeneous and thus their requests are heterogeneous, we also consider using a stochastic request arrival model for more efficiency in terms of resource provisioning. We model the problem as a mean-field game (MFG) in which the players are the VNFs competing over the MEC infrastructure resources. Precisely, we model the interaction between the different VNFs as a behavioral study shaped as an MFG. We prove that the MFG can be reduced to a discrete-time Markov decision process (MDP) model with continuous state and action spaces. Therefore, the solution to the MDP is also a solution to the MFG problem. To solve the MDP, we consequently leverage reinforcement learning (RL) tools to learn complex reward functions, policies, and forwarding dynamics through a guided cost learning paradigm and the actor-critic framework.

The list of the contributions of this paper is given in the following.
\begin{itemize}
    \item We propose an MFG-based behavioral study for the VNFs placement and chaining problem.
    \item We prove that the MFG can be reduced to an MDP through different steps of transition.
    \item We propose a RL-based algorithm to solve the MDP that is able to learn the VNFs placement and chaining policy through the actor-critic approach.
\end{itemize}

\begin{figure}[!h]
    \centering
    \includegraphics[width=.7\linewidth]{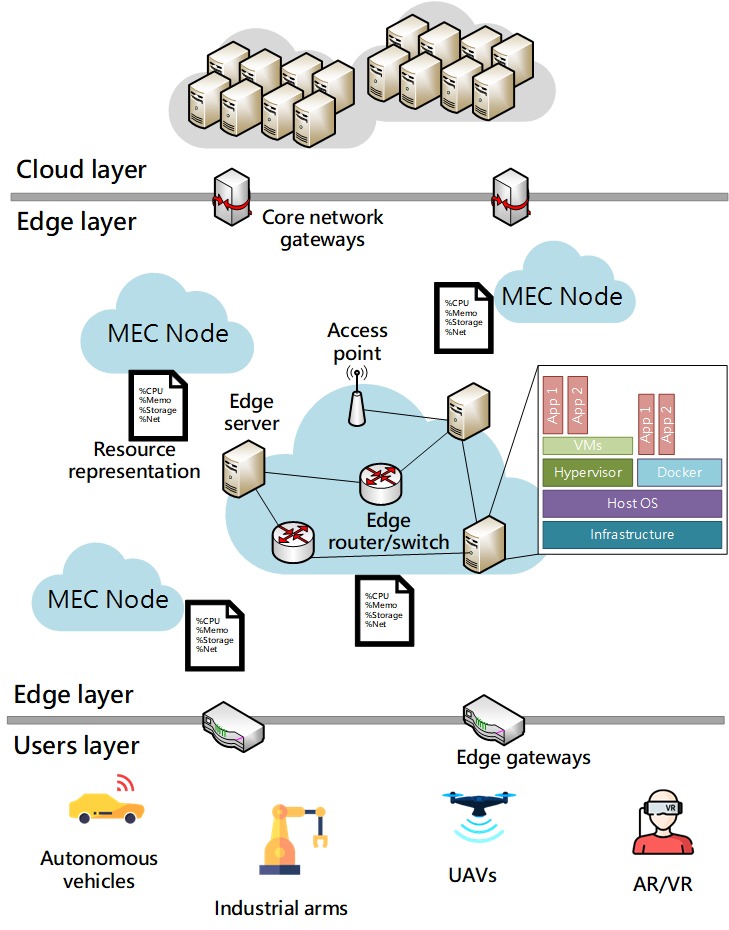}
    \caption{\label{fig:edge-arch} An illustration of the considered MEC-enabled SFC architecture.}
\end{figure}

The remainder of this paper is structured as follows. In Section II, we present the system model and formulate the problem of SFC resource provisioning. In Section III, we provide the details about the proposed solutions. The performance of the system is presented in Section IV. Last but not least, the related works are discussed in Section V. Finally, the paper is concluded in Section VI.

\begin{figure}[!h]
    \centering
    \includegraphics[width=.7\linewidth]{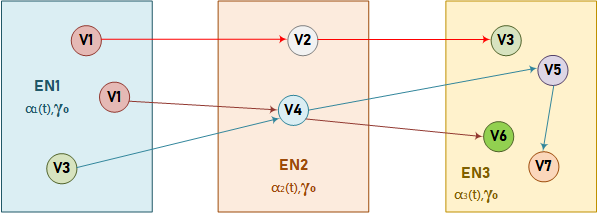}
    \caption{\label{fig:edge-nodes} An illustration of SFC with VNFs distributed over multiple MEC nodes.}
\end{figure}

\section{System Model}
In this section, we present the different components of the considered system model. We introduce first, the network architecture, then the computational and transmission model, the end-to-end (E2E) delay, and finally the problem formulation. In this paper, we consider a slotted system with $t \in \mathbb{N}$ being the time slots and $\mathbb{N}$ representing the set of natural numbers.
\subsection{Network Architecture}
Let us consider a MEC-enabled network as depicted in Fig. \ref{fig:edge-arch}, in which a set of users depicted on the bottom side of the figure request services from the edge layer. These services are being deployed as a sequence of VNFs distributed over different MEC nodes. Users can assigned through an assignment model such as in \cite{8647545, 9383093}, unlike the work in \cite{9524882}, the mobility was not considered in this work. Let $\mathcal{E}=\{e_1, e_2, \dots, e_N\}$ be the set of MEC nodes. Each MEC node is built up on a set of edge devices, each having a resource set of computing, storage, and transmission. Without loss of generality, we consider only the placement of the VNFs within a given MEC node regardless of the specific assignment between the VNFs and the devices of the corresponding MEC node. In addition, we consider the overall resource state of each MEC node with no regard to the specific amount of resource on each edge device. Such condition is met through the resource representation technique proposed in \cite{9326402,abouaomar2019resources}. Let $\alpha_i$ be the amount of available resources on each MEC node $i$. We consider the computation $c_i$ resource, the storage resource $s_i$ and the transmission resource $\omega_i$ that together define $\alpha_i$ as follows:
\begin{equation}
    \alpha_i=\left\langle c_i, s_i, \omega_i \right\rangle.
\end{equation}

Each MEC node hosts a set of VNFs $\mathcal{F}_i=\left\{f_{(1,i)}, f_{(2,i)}, \dots, f_{(J,i)}\right\}$. Each VNF $j$ requires an amount of resource $\hat{\alpha}_j$, in terms of computation, storage, and transmission, to perform its task, defined as:
\begin{equation}
    \hat{\alpha}_j=\left\langle \hat{c_j}, \hat{s_j}, \hat{\omega_j} \right\rangle.
\end{equation}
In order to have a better view on the placement of the VNFs, we consider the binary variable $x_{(i,j)}$ to denote the VNF-MEC node assignment, i.e., $x_{(i,j)}=1$ if and only if VNF $j$ is instantiated in MEC node $i$.

Let us consider a set of services $\mathcal{S}=\left\{s_1, s_2, \dots, s_{M}\right\}$, where each service $s_k$ is the composition of a set of VNFs, and it is defined as:
\begin{equation}
    s_k=\bigcup_{\substack{i\in \mathcal{E} \\ j\in{J}_{(k,i)}}} \left\{f_{(j,i)}\right\},
\end{equation}
and $J_{(k,i)}$ is the set of VNF used by service $k$ in MEC node $i$ and $f_{(j,i)}$ is the VNF $j$ on MEC node $i$ taking part of the SFC $s_k$. We consider that the VNFs are sorted in a way that the first element of $s_k$ is the ingress and the last element is the egress. Fig.~\ref{fig:edge-nodes} gives an example where $s_1=\{f_{(1,1)},f_{(2,2)},f_{(3,3)}\}$, $s_2=\{f_{(1,1)},f_{(4,2)},f_{(6,3)}\}$ and $s_3=\{f_{(1,1)},f_{(4,2)},f_{(5,3)},f_{(7,3)}\}$. We consider the binary variable $x_{(i,j)}^k$ to denote the VNF-MEC-SFC assignment, defined as follows:
\begin{equation}
    \label{eq:vnf-mec-sfc}
    x_{(i,j)}^{k}=\begin{cases}
        1 \text{ if the VNF $j$ on MEC node $i$ in SFC $k$}\\
        0 \text{ otherwise}.
    \end{cases}
\end{equation}

We consider that the MEC nodes communicate and a portion of the channel is allocated to the VNFs in order to forward their packets to VNFs hosted on other MEC nodes. Let $L_{(i,i')}$ be the link capacity between the MEC nodes $i$ and $i'$ and we let $l_{(i,j)}^{(i',j')}$ be the portion of $L_{(i,i')}$ allocated for the VNF $j$ on MEC node $i$ to forward the packets to the VNF $j'$ on MEC node $i'$.
We consider that the channel access is allocated using orthogonal frequency division multiple access (OFDMA). For clearness sake, in the rest of the paper, we consider that the index $i$ denotes the MEC nodes, $k$ denotes the SFCs and $j$ denotes the VNFs.

\subsection{Processing Model}
Before discussing the computational model, it is necessary to define the request format and the request arrival process. We define the user's $u$ request for service $k$ as follows:
\begin{equation}
    \rho_u^k=\left\langle s_k, \beta_u, T_{out}\right\rangle,
\end{equation}
where $s_k$ is the requested service, $\beta_u$ is the packet size, and $T_{out}$ the timeout for the request. The computational model is defined as the number of time slots required to process a request having the packet size $\beta_u$. Assuming that the packet passes through a set of VNFs, the processing time is defined as the sum of the computational times spent by the packet on each MEC node. Let $\delta_{(i,j,u,k)}$ be the computational time spent by VNF $j$ to process the request $\rho_u^k$ on MEC node $i$, defined as follows \cite{9326402}:
\begin{equation}
    \delta_{(i,j,u,k)}^p=x_{(i,j)}^{k}\frac{\hat{c_j}\beta_u}{\alpha_i}
\end{equation}
where $\alpha_i$ is the processing capacity of the MEC node $i$ ($p$ stands for processing). 
The overall computational time experienced by user $u$ requesting service $k$ is given as follows:
\begin{equation}
    \delta_{k,u}^{p}=\sum_{i\in\mathcal{E}} \sum_{j\in\mathcal{F}_i} \delta_{(i,j,u,k)}^{p}.
\end{equation}
When user $u$ request service $s_k$, let $T_{(i,i',j,j')}^{u,k}$ be the transmission time experienced when forwarding packets from VNF $j$ of MEC node $i$ to VNF $j'$ of MEC node $i'$. Consequently, the overall transmission delay is defined as the sum of the transmission times between the different VNFs of the requested SFC in the MEC nodes. The transmission time $T_{(i,i',j,j')}^{u,k}$ is defined as follows:
\begin{equation}
    T_{(i,i',j,j')}^{u,k}=\frac{\hat{\omega_j}\beta_u}{l_{(i,j)}^{(i',j')}}x^k_{(i,j)}x^k_{(i',j')},i\neq i',j\neq j',
\end{equation}
where VNF $j$ is associated to MEC node $i$ and VNF $j'$ is associated to MEC node $i'$. For $i=i'$, the transmission delay is defined to be zero.

The total transmission time for user $u$ requesting service $s_k$ is given as:
\begin{equation}
    \delta_{k,u}^{t}=\sum_{i\in\mathcal{E}}\sum_{i'\in\mathcal{E}\backslash\{i'\}}\sum_{j\in\mathcal{F}_i}\sum_{j'\in\mathcal{F}_{i'}}T_{(i,i',j,j')}^{u,k},
\end{equation}
where $t$ stands for transmission. Finally, the overall delay experienced by a request $\rho_u^k$ is given as follows:
\begin{equation}
    \delta_{(k,u)}=\delta_{k,u}^{p} + \delta_{k,u}^{t}.
\end{equation}
\subsection{Problem Formulation}
In this paper, we investigate the problem of SFC resource provisioning within MEC in 5G and beyond networks by placing and chaining the different VNFs in order to offer low overall delay. The problem is defined as follows:

\begin{mini!}[2]
  {}{\sum_{u\in \mathcal{U}}\sum_{k=1}^M\delta_{(k,u)}}
  {\label{eq:delay}}{}
  \addConstraint{\sum_{\substack{i\in \mathcal{E}}}x_{(i,j)}^k}{\leq1,j\in\mathcal{F}_i,s_k\in\mathcal{S}}
  \addConstraint{
    \sum_{\substack{j\in\mathcal{F}_i}}\sum_{s_k\in\mathcal{S}}\hat{\alpha}_{j}x_{(i,j)}^k
   }{\leq\alpha_i,i\in\mathcal{E}}
\end{mini!}
Where constraint in (\ref{eq:delay}b) guarantees the resource consumption, which means that the overall resource being allocated to all VNFs in MEC node $i$ should not exceed the total available resources. Constraint in (\ref{eq:delay}c) guarantees the packet's flow conservation.

\section{Mean-Field Resource Allocation}

\subsection{Mean-Field Game Formulation}
In order to solve the problem of SFC resource provisioning, we model it as a mean-field game (MFG). In the sequel, we define the MFG framework requirements. The SFC is modeled as an oriented graph $G_k=\left\langle L_k,\mathcal{F}_k\right\rangle$, where $\mathcal{F}_k$ is the set of VNFs composing the SFC (\ie the vertices or the states) and $L_k$ the set of virtual links (\ie the edges). The graph $G_k$ is constructed over the MEC nodes as illustrated in Fig. \ref{fig:edge-nodes}, and the edges are allocated over the horizontal link existing between the MEC nodes. Without loss of generality, we focus on service $s_k$. Let $\theta_i^j(t)$ be the density (\ie the number) of the VNF $j$ instantiated on MEC node $i$ at the time slot $t$. Let $\theta(t)$ be the set of the densities defined as:
\begin{equation}
    \theta(t)=\left\{\theta_i^j(t) \text{ with } i\in \mathcal{E}, j\in\mathcal{F}_i\right\}.
\end{equation}

We consider that VNFs are placed and chained in each time slot following a dynamic generated by the matrix $\mathcal{P}(t)$ that is a stochastic matrix, in which its rows and columns are transition probabilities between the different states (the VNFs). For example, the value $P_{jj'}(t)$ represents the transition probability between VNF $j$ and VNF $j'$. For each state $j$, we define a value function $V_j(t)$ for each distribution of the VNFs at time slot $t$, and we define a reward function $r_j(\cdot)$ as the payoff for the VNF $j$. The reward depends on the distribution $\theta(t)$ and the row vector $\mathcal{P}_j(t)$. We consider a continuous-time MFG, and for each state $j$, we define the Hamilton-Jacobi-Bellman (HJB) equation as follows \cite{yang2017learning}:
\begin{equation}
    \label{eq:hjb}
    V_j(t)=\max_{\mathcal{P}_j(t)}\left\{r_j\left(\theta(t), \mathcal{P}_j(t)\right)+\sum_{\substack{j'\in\mathcal{F}_i \\ j\neq j'}}P_{jj'}(t)V_{j}(t+1)\right\}.
\end{equation}
The Fokker-Plank-Kolmogorov (FPK) equation is defined as follows \cite{yang2017learning}:
\begin{equation}
    \label{eq:fpk}
    \theta_{j}(t+1)=\sum_{j'\in\mathcal{F}_i}P_{jj'}(t)\theta_{j'}(t),
\end{equation}
where $\theta_j(t)$ represents the distribution of VNF $j$.

The HJB and FPK equations given in Eq. (\ref{eq:hjb}) and Eq. (\ref{eq:fpk}) are also called the backward and forward functions respectively. The backward function enables computing the optimal placement for each VNF, and the forward function represents the evolution of the mean-field term for the VNFs.

\subsubsection{VNFs Distribution}
The VNFs distribution is represented as a discrete probability distribution of the set of VNFs. We denote the distribution of VNF $j$ as $\theta_j(t)$, which represents the instances at time $t$ of the set of VNF $j$.
\subsubsection{Reward}
The reward is received by the VNF $j$ instance choosing a given action $\mathcal{P}_j(t)$ at the time $t$. We also assume that the reward for a set of VNFs $j$ depends only on the actions taken in time slot $t$, and does not affect the other instance of VNFs $j'\in \mathcal{F}_i$. We define the reward as follows:
\begin{equation}
    \label{eq:reward}
    r_j(\theta(t),\mathcal{P}_j(t))=\sum_{j'\in \mathcal{F}_i}P_{jj'}(t)r_{jj'}(t),
\end{equation}
where $r_{jj'}(t)$ is the reward received after moving from VNF $j$ to VNF $j'$.

\subsubsection{Transition Probabilities}
The transition probability is represented by the matrix $\mathcal{P}_j(t)$ defined in the previous section. It represents the probability on which the VNF $j$ choose to change its current action and thus transitions to a new state. In addition, $\mathcal{P}(t)$ is the FPK generator, and we keep the same notation in Eq. (\ref{eq:fpk}) to denote the forward equation.
\subsubsection{The Average Reward}
The average reward is the value of the reward received by the VNF $j$ instances when a given distribution $\theta(t)$ and the action $\mathcal{P}(t)$ are taken. Also, the average reward is the value that each VNF is willing to maximize. We define the average reward as follows:
\begin{equation}
    \label{eq:mu}
    \mu_j(\theta(t), \mathcal{P}(t), V_j(t))=\sum_{j\in \mathcal{F}_i}P_{jj'}(t)r_{jj'}(\theta(t),\mathcal{P}(t))+V_{j'}(t).
\end{equation}

\begin{figure*}[!h]
  \centering
  
\vspace{.2in}
  \begin{subfigure}{.3\linewidth}
    \centering
    \includegraphics[width = \linewidth]{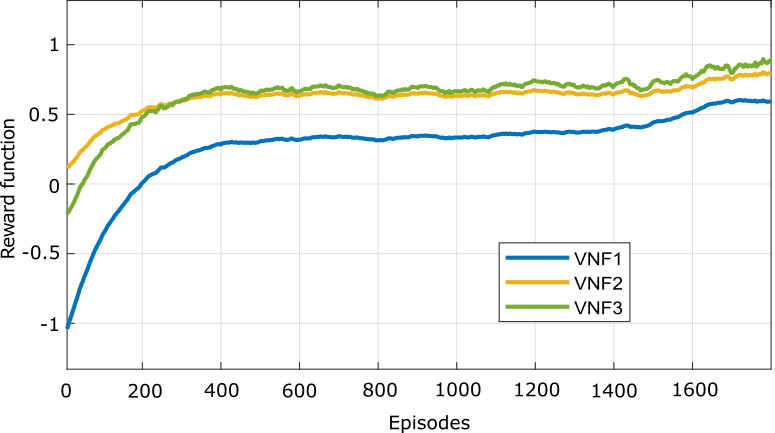}
    \caption{SFC-1}
  \end{subfigure}%
  \hspace{1em}
  \begin{subfigure}{.3\linewidth}
    \centering
    \includegraphics[width = \linewidth]{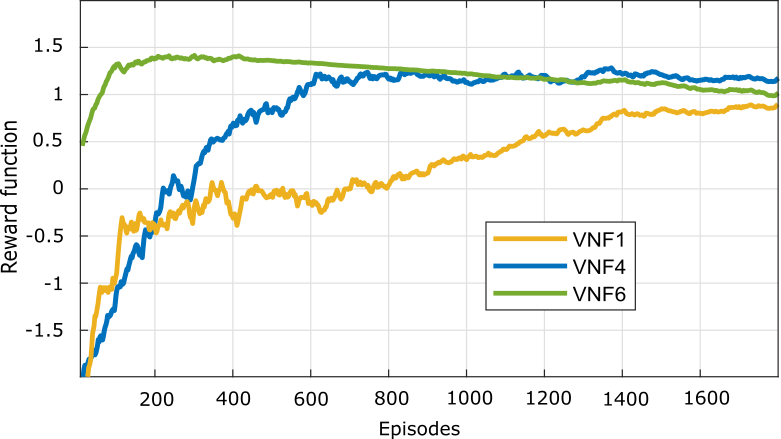}
    \caption{SFC-2}
  \end{subfigure}%
  \hspace{1em}
  \begin{subfigure}{.3\linewidth}
    \centering
    \includegraphics[width = \linewidth]{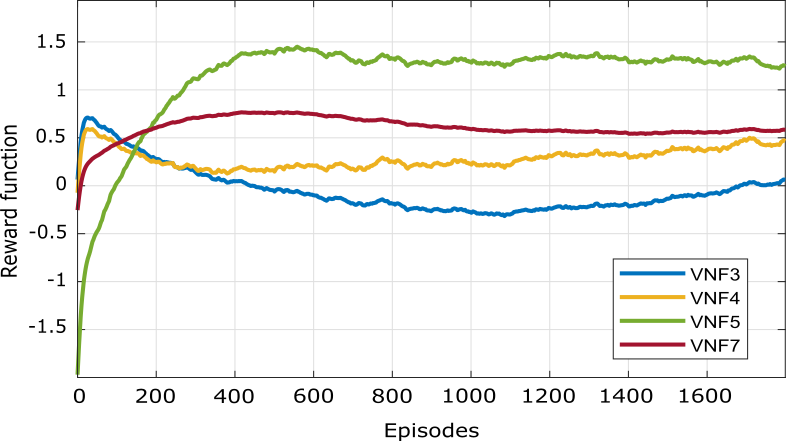}
    \caption{SFC-3}
  \end{subfigure}
  \caption{Reward function of the proposed learning approach for the different considered scenarios.}
  \label{fig:rewards}
\end{figure*}

In order to prove the existence of a Nash equilibrium we prove the existence of a Nash maximizer matrix over the transition probabilities matrix. Let $P(\mathcal{P}, j, q)$ equal to the matrix $P$ with the row $j$ replaced by $q$. 
Therefore, a Nash maximizer is defined as $\mu_j(\theta, \mathcal{P}, V)\geq\mu_j(\theta, P(\mathcal{P}, j, q), V)$, where $V$ and $\mu$ are the value function and the average reward, respectively \cite{yang2017learning}.
At Nash equilibrium, the VNFs have no intention to change their actions in order to increase the reward but instead can choose actions as in $q$. Therefore, we can express the value function as follows:
\begin{equation}
    \label{eq:new_v}
    V_j(t)=\max_{q}\left\{\sum_{j\in\mathcal{F}_i}q_j\left[r_j(\theta(t),P(\mathcal{P}(t), j, q))+V_j(t+1)\right]\right\},
\end{equation}
where $q_j$ is an element of the vector $q$. The solution to the MFG is given by the set of $\langle\theta(t), V(t)\rangle$ with consideration to the requirements in Eq. (\ref{eq:mu}) and Eq. (\ref{eq:new_v}).

\subsection{The MFG Reduction to MDP}
We reduce the proposed MFG to the MDP model defined by the following parameters.
\subsubsection{States} The states are defined in our case as the distribution of the VNFs within the same MEC node at the time slot $t$. We keep the same notation, $\theta_j(t)$, for the density of VNF $j$ at time slot $t$.
\subsubsection{Actions} The actions are the matrices of the transition probabilities as given with $\mathcal{P}(t)=[P_{jj'}(t)]$. We denote by $\mathcal{P}_i$ the actions that a VNF can take on a MEC node.
\subsubsection{Reward} The reward is defined as the payoff received by the VNFs within the same node. We denote it by $R(\theta,\mathcal{P})$ and it is given as follows:
\begin{equation}
    R(\theta_j(t),\mathcal{P}_i(t))=\sum_{j \in s_k}\sum_{j'\in \mathcal{F}_i}P_{jj'}(t)r_{jj'}(t)
    \label{eq:mdpr}
\end{equation}

\subsubsection{Transition matrix}
The transition matrix is equivalent to the FPK equation in Eq. (\ref{eq:fpk}), and given as,
\[
    \theta_{j}(t+1)=\sum_{j'\in\mathcal{F}_i}P_{jj'}(t)\theta_{j'}(t).
\]

The solution's value function of a MFG over the graph $G_k$ defined by (i) $V_j(t)$ in Eq. (\ref{eq:new_v}), which is also the HJB equation and (ii) the forward equation in Eq. (\ref{eq:fpk}), which represents the FPK equation, is a solution to the Bellman equation of an MDP equivalent to the one defined through the previous definitions \cite{yang2017learning}. In addition, if we consider a initial state $\theta_j(0)$ the optimal policy of an MDP is equivalent to the FPK equation.

\begin{algorithm}[!t]

\SetAlgoLined
\SetKwInOut{Initialization}{Initialization}
    \KwIn{$V_j(0)$, $Episodes$, $\theta_j(0)$}
    \Initialization {Initialize $\mathcal{P}_j$}
    \For{$e \gets 0$ \text{ to } $Episodes$}{
        Generate initial distribution of VNFs\;
        Generate samples from $\mathcal{P}_{j'}, j\neq j'$\;
        Compute the average reward $\mu_j(\theta_j,\mathcal{P}_j), V_j)$ from Eq. (\ref{eq:mu})\;
        Initialize the policy vectors\;
        \For{$t\gets0$ \text{ to  } $Samples$}{
            Generate action based on $P$\;
            Compute reward $r_j(\theta(t),\mathcal{P}_j(t))$ from Eq. (\ref{eq:mdpr})\;
            Update the reward from Eq. (\ref{eq:mu})\;
            Update the policy from Eq. (\ref{eq:new_v})\;
        }
    }
    \Return final reward\;
    \caption{Adopted Actor-Critic-based reinforcement learning inspired from \cite{hu1998multiagent} and \cite{pmlr-v48-finn16}}
\end{algorithm}

Algorithm 1 provides the steps adopted to find the optimal policy for VNFs placement and chaining based on the reinforcement learning paradigm, especially, the actor-critic RL framework. The algorithm is inspired from \cite{hu1998multiagent} and \cite{pmlr-v48-finn16} in which the actors (VNFs in our case) act on the critics provided by the infrastructure controller to adjust their behavior. In our case, the critics can guide the actors in a manner to maximize their value function, and therefore, be placed and chained adequately, allowing the minimization of the overall delay.
In simple words, at each episode, we generate a distribution of the VNFs, we sample probabilities transitions $P_j$, and we compute the average reward $\mu_j$ for that distribution. We initialize then the policy vector and we start the learning phase through a number of training samples. At each training sample, we generate a set of actions, and we evaluate the impact of these actions on the system through computing the reward according to Eq. (\ref{eq:mdpr}). The algorithm then updates the overall reward and the policy through maximizing the value function in Eq. (\ref{eq:new_v}).

\begin{figure*}[!t]
  \centering
\vspace{.2in}
  \begin{subfigure}{.3\linewidth}
    \centering
    \includegraphics[width = \linewidth]{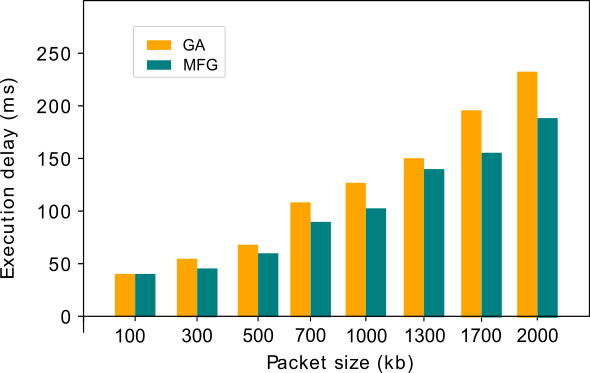}
    \caption{SFC-1}
  \end{subfigure}%
  \hspace{1em}
  \begin{subfigure}{.3\linewidth}
    \centering
    \includegraphics[width = \linewidth]{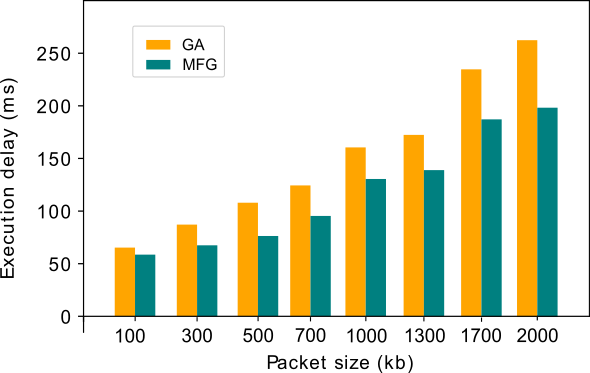}
    \caption{SFC-2}
  \end{subfigure}%
  \hspace{1em}
  \begin{subfigure}{.3\linewidth}
    \centering
    \includegraphics[width = \linewidth]{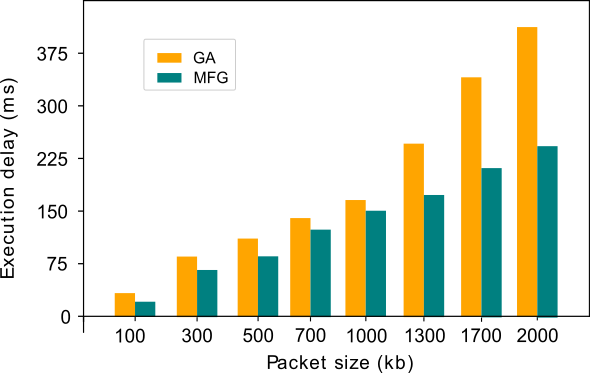}
    \caption{SFC-3}
  \end{subfigure}
  \caption{Average delay for under different scenarios.}
  \label{fig:delays}
\end{figure*}

\section{Experimentation and Simulations}

In this section we provide the simulation results for the proposed approach. We compare the proposed learning approach to a genetic algorithm (GA).

We consider the architecture provided in Fig. \ref{fig:edge-nodes} consisting in $3$ MEC nodes, $3$ services, and each service consists in a number of VNFs given as follow:
\begin{itemize}
    \item SFC-1 :\{VNF-1, VNF-2, VNF-3\}
    \item SFC-2 :\{VNF-1, VNF-4, VNF-6\}
    \item SFC-3 :\{VNF-3, VNF-4, VNF-5, VNF-7\}
\end{itemize}

We consider a request generator that generates requests with a size varying between $100Ko$ and $2Mo$. We also performed the training over $2000$ episodes and we consider a resource requirement for the different VNFs between $10\%$ and $70\%$ of the available resources.

Fig. \ref{fig:rewards} shows the different rewards values for the different VNFs composing the considered scenarios. Note here that the reward depends on the placement of the VNFs within a given node, and on the available resource at each node. In the MEC node 1, VNF-1 exists twice and taking part of two different SFCs, and it is shown in Fig. \ref{fig:rewards}(a) and \ref{fig:rewards}(b) that they almost converge to the same reward value. However, in the case when a single VNF is embedded within different SFCs, which is the case of VNF-4, its reward converges to two different values, which impacts also the delay as we detailed in Fig. \ref{fig:delays}(b) and \ref{fig:delays}(c). Also, the reward is impacted by the number of VNFs within the same node, which is the case of MEC node 2.

Fig. \ref{fig:delays} illustrate the overall execution delays of the considered scenarios as a function of the packet size. We compared the obtained results with a benchmark of a literature solution that leverage GA to reduce the overall SFC execution delay. The GA method relay on solving the optimization problem through applying crossover and mutation operations. Each potential solution is encoded as a chromosome and the quality of each solution is evaluated using a fitness function. The GA method begins its operations with a initial set of possible solutions, and as things progress the solutions are being tuned to give birth to new generations of enhanced quality through a crossover operation then generate a new set of solutions from selected parents then mutate them to enhance the chromosome.
In the illustrated results, it is clear that the proposed RL-based solution outperformed the benchmarked GA-based solution with different values of the packet sizes. The obtained results also shows the effect of the SFC size (the number of VNFs composing it) which is clear from the difference in delays obtained in Fig. \ref{fig:delays}(a-b) and \ref{fig:delays}(c) which got higher when adopting a GA-based placement and chaining scheme (250ms in (a) and (b), and ~375ms in (c)), while slightly got high when considering our MFG-RL based solution (200ms in (a) and (b), and ~225ms in (c)). It is also due to the fact that the transmission delay between VNFs within the same nodes is considered to be negligible as detailed in the system model section.

\section{Related Works}
The work in \cite{8647858} proposed a meta-heuristic-based scheme for VNF placement within a MEC environment under constraints of minimizing latency and maximizing the service availability. The authors proposed a GA to cope with the complexity of the problem and studied the performance compared to a CPLEX implementation of their model. However, some aspects related to the dynamic nature of the model were not considered in their proposed solution, namely the request arrival and diversity of the services. The work in \cite{9062531}, proposed an integer linear programming formulation to the problem of SFC resource allocation, which is NP-hard. To cope with the high complexity of the problem, the authors proposed a greedy algorithm that leverages the different features of the considered infrastructure. In \cite{9139632} authors investigated the problem of VNF placement at the edge with the aim to maximize the users' benefits. The authors proposed two formulations to study the problem of VNF placement through the investigation of the capacitated MEC allocation problem. First, the authors proposed an integer-linear programming formulation and in the second formulation, they studied the VNFs as a coverage problem. The solutions are based on a sub-modular optimization to approximate the optimal solution. The work in \cite{sallam2019joint} proposed a joint VNF-placement and resource allocation scheme to maximize the network flows under budget and capacity constraints of the placement infrastructure. Such formulation is NP-hard and non-sub-modular, hence, the authors proposed a relaxation method to make the placement problem sub-modular. 

In this paper, we consider a more dynamic model related to the request arrival, the heterogeneity of the VNFs, and the heterogeneity of resource requirements of the different VNFs. In addition, we propose using an efficient RL-based approach that enables the system to learn different placement and chaining policies based on the actor-critic approach.

\section{Conclusion}
In this paper, we investigated the problem of service function chaining (SFC) resource provisioning while considering constraints on the placement and the chaining of virtual network functions (VNFs) within a multi-access edge computing (MEC) infrastructure. We aimed at reducing the operational cost in terms of delay while placing and chaining the VNFs. We proposed a behavioral study for the VNFs that need to be placed and chained accordingly to offer reduced delay. We proposed a mean-field game (MFG) framework to model the behavior of the VNFs. In addition, we reduced the MFG formulation of the problem to a Markov decision process (MDP) in order to reach the optimal resource provisioning policy with no need for system control parameters as in classic theoretical game models. Specifically, we leveraged the reinforcement learning (RL) approach using an actor-critic model to make the MEC nodes learn complex reward functions, policies, and the forwards dynamic. 

In future work, we will leverage the results to model the service popularity and service deployment prediction for better resource management, hence, a better quality of service. Additionally, the introduction of novel concepts such as C-V2X \cite{9497103}, and network slicing \cite{9318243}.

\section*{Acknowledgment}
The authors would like to thank the Natural Sciences and Engineering Research Council of Canada, for the financial support of this research.

\bibliography{references} 
\bibliographystyle{IEEEtran}

\end{document}